# Layer-dependent superconductivity in iron-based superconductors

Ke Meng[1,2,*], Xu Zhang[1,*], Boqin Song[1,*], Baizhuo Li[3,*], Xiangming Kong[1], Sicheng Huang[1], Xiaofan Yang[1], Xiaobo Jin[1], Yiyuan Wu[1], Jiaying Nie[1], Guanghan Cao[3,4,✉], and Shiyan Li[1,2,4,✉]

[1]*State Key Laboratory of Surface Physics, Department of Physics, Fudan University, Shanghai 200438, China*

[2]*Shanghai Research Center for Quantum Sciences, Shanghai 201315, China*

[3]*School of Physics, Zhejiang University, Hangzhou 310058, China*

[4]*Collaborative Innovation Center of Advanced Microstructures, Nanjing 210093, China*

Corresponding author. Email: shiyan_li@fudan.edu.cn (S.Y.L.); ghcao@zju.edu.cn (G.H.C.)

**Abstract**

**The Hohenberg-Mermin-Wagner theorem states that a two-dimensional system cannot spontaneously break a continuous symmetry at finite temperature. This is supported by the observation of layer-dependent superconductivity in the quasi-two-dimensional superconductor NbSe$_2$, in which the superconducting transition temperature ($T_c$) is reduced by about 60% in the monolayer limit. However, for the extremely anisotropic copper-based high-$T_c$ superconductor Bi$_2$Sr$_2$CaCu$_2$O$_{8+\delta}$ (Bi-2212), the $T_c$ of the monolayer is almost identical to that of its bulk counterpart. To clarify the effect of dimensionality on superconductivity, here we successfully fabricate ultrathin flakes of CsCa$_2$Fe$_4$As$_4$F$_2$, a highly anisotropic iron-based high-$T_c$ superconductor, down to monolayer. The monolayer flake exhibits the highest $T_c$ of 24 K (after tuning to the optimal doping by ionic liquid gating), which is about 20% lower than that of the bulk crystal. We also fabricate ultrathin flakes of CaKFe$_4$As$_4$, another iron-based superconductor with much smaller anisotropy. The $T_c$ of the 3-layer flake decreases by 46%, showing a more pronounced**

**dimensional effect than that of CsCa$_2$Fe$_4$As$_4$F$_2$. By carefully examining their anisotropy and the *c*-axis coherence length, we reveal the general trend and empirical law of the layer-dependent superconductivity in these quasi-two-dimensional superconductors. From this, the $T_c$ of a new monolayer superconductor can be extrapolated.**

**Introduction**

When a three-dimensional (3D) crystalline superconductor is thinned towards the two-dimensional (2D) limit, a crucial question is whether the superconductivity in the monolayer differs from that in the bulk crystal. Based on the Hohenberg-Mermin-Wagner theory, superconductivity, as a long-range order, is strongly suppressed in systems with reduced dimensions[1-4]. This is supported by the experimental results of the quasi-2D superconductor NbSe$_2$, in which the superconducting transition temperature ($T_c$) decreases significantly as the thickness approaches the monolayer limit[5]. However, the monolayer high-$T_c$ cuprate superconductor Bi$_2$Sr$_2$CaCu$_2$O$_{8+\delta}$ (Bi-2212) exhibits a $T_c$ as high as that of the bulk single crystal[6], which seems to contradict the theory. It implies that Bi-2212 is essentially a 2D superconductor.

The 2D nature of Bi-2212 is apparently related to its extremely high anisotropy, here defined as the ratio of *c*-axis and *ab*-plane resistivity $\rho_c/\rho_{ab}$, which is in the order of $10^5$ (refs. [7,8]). Its 2D nature also manifests in the extremely short superconducting coherence length along the *c* axis ($\xi_c \leq 0.6$ Å)[9], suggesting that the superconducting Cooper pairs of Bi-2212 is basically confined in the CuO$_2$ planes. In contrast, NbSe$_2$ has a much smaller anisotropy ($\rho_c/\rho_{ab} \sim 10^2$)[10] and a much longer $\xi_c$ (~ 23 Å)[11]. The distinct superconducting behaviors in the 2D limit for NbSe$_2$ and Bi-2212 raise the question in general how the dimensionality affects the superconductivity in quais-2D superconductors. This will have significant effect on the application of van der Waals superconductor thin flakes or heterostructures[12].

To clarify this issue, we focus on two related iron-based superconductors: the 12442 system ($A$Ca$_2$Fe$_4$As$_4$F$_2$, $A$ = K, Rb, Cs)[13,14] and the 1144 system (Ca$A$Fe$_4$As$_4$, $A$

= K, Rb, Cs)[15]. The 12442 system is developed from the 1144 system by replacement of the Ca layers with $Ca_2F_2$ layers, as shown in Fig. 1a. Both systems share the same intrinsic hole concentration near optimal doping and the same range of $T_c$ (from 30 to 35 K), but their anisotropy differs significantly[16,17]. Like Bi-2212, the 12442 system has the superconducting double $Fe_2As_2$ sheets separated by insulating layers. Such a layered structure leads to its highest anisotropy (~ 3150, ref.[16]) among all iron-based superconductors. In contrast, the anisotropy of 1144 system superconductor is much smaller (~ 10, ref.[17]), indicating a much stronger interlayer coupling. These features make them ideal compounds for exploring the role of the dimensional effect on quasi-2D superconductors.

In this work, we successfully fabricate large-area thin flakes of $CsCa_2Fe_4As_4F_2$ (Cs-12442) down to monolayer by the $Al_2O_3$-assisted exfoliation method and study its layer-dependent superconductivity through electrical transport measurements. As the flake thickness decreases, the $T_c$ initially maintains unchanged (about 30 K) until 4 layers, and then decreases 20% to 24 K for the monolayer (after modulated by ionic liquid gating to minimize the effect of doping change from the fabrication process). The signature of the Berezinskii-Kosterlitz-Thouless (BKT) transition indicates a 2D feature of the superconductivity in the monolayer Cs-12442. Furthermore, thin flakes of $CaKFe_4As_4$ were also fabricated down to 3 layers, and a reduction of $T_c$ by 14%, 22%, and 46% was observed in the 5-layer, 4-layer, and 3-layer samples, respectively. We carefully examine the anisotropy and the $c$-axis coherence length of Bi-2212, Cs-12442, $NbSe_2$, and $CaKFe_4As_4$, and figure out the general trend and empirical law of dimensional effect on the superconductivity in the 2D limit for layered superconductors.

## Results

**Fabricating Cs-12442 thin flakes**

As plotted in Fig. 1a, Cs-12442 has a layered structure with double $Fe_2As_2$ layers separated by $Ca_2F_2$ layers. Notably, one unit cell of Cs-12442 contains two electronically equivalent $Cs-Fe_2As_2-Ca_2F_2-Fe_2As_2-Cs$ half-unit cell. Experimentally, obtaining large-area monolayer flakes via conventional mechanical exfoliation in this system turns out to be quite difficult[18], since the in-plane bonding is not strong enough and the thin flakes will be fragmented during repetitive exfoliation processes. To overcome the challenge, we apply the $Al_2O_3$-assisted exfoliation method, which is an efficient way to isolate monolayers from bulk single crystals that are otherwise difficult to exfoliate by conventional methods, such as $Fe_3GeTe_2$[19], $MnBi_2Te_4$[20], $FeSe$[21], $CsV_3Sb_5$[22] and $Fe_5GeTe_2$[23]. Figure 1b shows the exfoliated Cs-12442 thin flakes with different layers. Here, two criteria are combined together to precisely determine the layer numbers: the optical transmittance and the atomic force microscope (AFM) topography. The optical transmittance of various numbers of layers follows the Beer-Lambert law (Fig. 1d), which enables us to identify different layers[19-22]. Meanwhile, the height difference between a contiguous number of layers extracted from AFM topography (Fig. 1c) is approximately 1.6 nm (Fig. 1e), which is consistent with the thickness of a half-unit cell of Cs-12442 (ref.[14]). Therefore, the number of layers can be accurately determined. Here, one half-unit cell of Cs-12442 is defined as one "monolayer" (1L).

**Layer-dependent superconductivity in Cs-12442**

The four-probe method with cold-welded indium electrodes or evaporated Cr/Au electrodes (device setup shown in Fig. 2a, b) is applied to measure the sample resistance. Figure 2c shows the temperature dependence of the normalized resistance ($R/R_{250\,K}$) for the Cs-12442 bulk single crystal and monolayer. The $R$ - $T$ curve of both samples shows hump-like anomaly around 150 K and linear behavior before the superconducting transition, which is consistent with previous reports[14,24]. More detailed normalized resistance $R/R_{50\,K}$ as a function of temperature for the bulk single crystal and thin flakes with different layer numbers are displayed in Fig. 2d. The $T_c$ (defined as the temperature where the differential of $R$ - $T$ curve reaches the maximum) of the bulk Cs-12442 is 30.5 K, and gradually decreases from 30.1 K to 19.6 K with the thickness reduced from 4 layers towards monolayer. It should be noted that, the superconducting transition width $\Delta T_c$ (defined as the width between 10% $R_n$ and 90% $R_n$, where $R_n$ is the resistance of normal state before superconducting transition) of the bulk ($\Delta T_c$ = 1.05 K) and the monolayer sample ($\Delta T_c$ = 1.18 K) are approximately the same, indicating the high quality and uniformity of the monolayer sample.

**2D superconductivity in monolayer Cs-12442**

The BKT transition, which can be revealed from $I$ - $V$ characteristic curves, is regarded as evidence for 2D superconductivity[25]. In this scenario, the zero-resistance state only emerges when the vortex and anti-vortex are bounded into pairs below the so-called BKT transition temperature ($T_{BKT}$)[26], resulting in a $V \sim I^\alpha$ behavior with $\alpha = 3$ at $T_{BKT}$. Figure 3a shows that the $I$-$V$ curves of the Cs-12442 monolayer (the same monolayer sample in Fig. 2d) at different temperatures. By fitting the $I$ - $V$ curves, the exponent $\alpha$ is deduced and shown in Fig. 3b. As the temperature decreases, the exponent deviates from 1 and reaches 3 at 19.0 K. Furthermore, the $R$ - $T$ curve follows a typical BKT-like behavior with $R(T) = R_0 \exp[-b(T/T_{BKT} - 1)^{-1/2}]$, when the temperature approaches $T_{BKT}$ (ref.[27]). Here, $R_0$ and $b$ are material parameters. A linear behavior should manifest in the plot of $(d\ln R/dT)^{-2/3}$ as a function of $T$. As shown in Fig. 3c, the extracted value of $T_{BKT}$ from the linear fitting is 19.1 K, which agrees well with the result of the $I$ - $V$ method. Therefore, $T_{BKT}$ obtained from both methods is very close to $T_c^{zero}$ (19.2 K, the temperature where the sample resistance reaches zero), demonstrating the 2D superconductivity in monolayer Cs-12442.

**Doping effect on the reduction of $T_c$**

As shown in Fig. 2d, a reduction of $T_c$ is observed in Cs-12442 thin flakes as the number of layers is lower than 4, especially in the monolayer. Before discussing the dimensional effect on the reduction of $T_c$, other extrinsic factors such as sample degradation and doping change need to be considered. Firstly, the ultrathin Cs-12442 flakes are very sensitive to outside environment. They can be easily oxidized, for example, during the preparation process in air. Therefore, our entire experimental process was completed in an Ar-filled glove box. The $T_c$ of monolayer sample can maintain essentially unchanged with a total exposure time 2~3 times longer than the typical sample preparation time (<

3 hours), as shown in Supplementary Fig. S2. Thus, the sample degradation in our experiments is negligible.

Secondly, it is well known that the $T_c$ of iron-based superconductors depends on the doping level, showing a superconducting dome[28-32]. In the 12442 system, the intrinsic hole doping originates from replacing alkali-earth metal atoms with alkali-metal atoms in the 122 system (e.g., $BaFe_2As_2$)[30]. According to the scanning tunneling microscope (STM) study on $KCa_2Fe_4As_4F_2$ (K-12442), most of the cleaved surfaces appear to be the alkali-metal atoms plane[33]. The bare Cs atoms on the cleaved surface can escape easily under the influence of the outside environment[33,34], since there is no chemical bond between Cs atoms and neighboring layers. Such a deficiency of surface Cs atoms has been observed in the exfoliated $CsV_3Sb_5$ thin flakes[22,35]. Therefore, there is a great possibility that the doping level may be changed during the exfoliation process, thereby affecting $T_c$. This kind of Cs deficiency on surfaces (as shown in Fig. 4a) has a negligible effect on the doping level for the bulk and thick flakes, but it has a conspicuous impact on the ultrathin flakes. Figure 4b shows the raw Hall data of Cs-12442 thin flakes with various numbers of layers at 35 K. The positive slop of the curves indicates that the hole carriers dominate the transport properties in all samples. The carrier density $n$ can be calculated by the formula $R_H = 1/ne$. Compared with the thick flake (360 nm), an increase in hole concentration for 4-layer and 2-layer Cs-12442 flakes is observed (here, the hole density in 360-nm thick flake is considered the same as in the bulk single crystal). In Fig. 4c, we plot the hole concentration of Cs-12442 ultrathin flakes on the phase diagram of $Ba_{1-x}K_xFe_2As_2$ (ref.[28]) to evaluate the doping effect on the reduction of $T_c$. The doping level calculated from Hall measurement is displayed by red circles, which clearly shows that Cs-12442 ultrathin flakes become more overdoped due to Cs deficiency on surfaces. This should be partly responsible for the reduction of $T_c$.

In monolayer Bi-2212, the $T_c$ can be tuned to the optimal value by changing the oxygen content through annealing in ozone[6]. Here for Cs-12442 ultrathin flakes, we employ ionic liquid gating to modulate the doping level, thus minimize the doping effect on $T_c$ reduction. Ionic liquid gating can efficiently modulate the carrier density in a large scale by inducing electrostatic or electrochemical effects[36-41]. This technique involves injecting small-sized ions (for instance, $H^+$) from the ionic liquid into the sample. Figure 4d shows the $R$ - $T$ curves of gated monolayer Cs-12442. An enhancement of $T_c$ is observed as the gating voltage ($V_g$), gating temperature ($T_g$) and gating duration time ($t_g$) increase. Similar enhancement in $T_c$ also occurs in 2-layer Cs-12442 (more details are shown in Supplementary Fig. S3 and S4). With applying a positive voltage, the small-sized $H^+$ ions are injected into the sample. The infused $H^+$ ions may fill the vacancies of surface Cs atoms, and induce electrons to the sample, so the hole concentration is reduced towards the optimal doping. The highest $T_c$ in the monolayer sample reaches 24 K at the condition of $V_g$ = 4.3 V, $T_g$ = 275 K and $t_g$ = 1 hour. With further changing the gating conditions (increasing $T_g$, $V_g$ and $t_g$), the $T_c$ is saturated. This saturation suggests that the doping level is likely tuned back to the flat part of the superconducting dome around optimal doping.

**Dimensional effect on the reduction of $T_c$**

In Fig. 4e, the evolution of $T_c$ with various layer numbers of Cs-12442 is plotted. For the monolayer and 2-layer, the highest $T_c$ of the gated samples can reach 24 and 27.5 K, respectively. Compared with the bulk single crystal, a reduction of $T_c$ still exists in the gated monolayer and 2-layer samples. Thus, the dimensional effect needs to be considered. To get more clues, we further fabricate thin flakes of CaKFe$_4$As$_4$ by the same exfoliation method. With decreasing flake thickness, a more significant reduction of $T_c$ (about 46% for the 3-layer CaKFe$_4$As$_4$) is observed, showing an even stronger dimensional effect (more details as shown in Supplementary Fig. S5).

Now we can discuss the dimensional effect on the $T_c$ reduction in layered superconductors. For quasi-2D superconductors, the 2D nature manifests in their anisotropy $\rho_c/\rho_{ab}$ and $c$-axis coherence length $\xi_c$. In Table 1, we list the $\rho_c/\rho_{ab}$, $\xi_c$, $\xi_c/d_{monolayer}$, $T_c$, and $-\Delta T_c/T_c^{bulk}$ for Bi-2212, Cs-12442, NbSe$_2$, CaKFe$_4$As$_4$, and their monolayers (3-layer for CaKFe$_4$As$_4$), to find out the general trend and empirical law. The anisotropy gradually decreases for these materials, from $1.5 \times 10^5$ in Bi-2212 to ~ 10 in CaKFe$_4$As$_4$. In Fig. 5a, the layer dependence of $T_c/T_c^{bulk}$ for them is plotted. It is evident that the material with smaller anisotropy shows a more pronounced $T_c$ reduction when approaching the 2D limit. Such a general trend can be seen more clearly in Fig. 5b, which plots the anisotropy dependence of $T_c$ reduction ($-\Delta T_c/T_c^{bulk}$) for these four materials with various numbers of layers.

The $c$-axis coherence length $\xi_c$ represents the distance that the Cooper pairs extend along the $c$ axis in quasi-2D superconductors, therefore it is a more important parameter to characterize the 2D nature of the superconductivity. For Bi-2212, $\xi_c \leq 0.6$ Å is only 1/25 of $d_{monolayer}$, the thickness of its monolayer[6,9]. This means that the Cooper pairs are basically confined in the $CuO_2$ planes, which can reasonably explain the absence of $T_c$ reduction in monolayer Bi-2212 (ref.[6]). For Cs-12442, although its $\xi_c$ (3.6 Å, ref.[16]) is less than $d_{monolayer}$ (about 1/5 of $d_{monolayer}$), this value is about 6 times larger than that of Bi-2212, indicating that the Cooper pairs likely extend outside the FeAs layers. Therefore, Cs-12442 shows a weak dimensional effect with 20% reduction of $T_c$ in the monolayer. While for $NbSe_2$ and $CaKFe_4As_4$, $\xi_c$ is larger than or comparable with $d_{monolayer}$[11,17]. The much-extended Cooper pairs along the $c$ axis in both materials results in a pronounced $T_c$ reduction as their thickness approaching the monolayer limit, which means a strong dimensional effect on superconductivity.

In Fig. 5c, the $T_c$ reduction vs. $\xi_c/d_{monolayer}$ is plotted for Bi-2212, Cs-12442, and $NbSe_2$. The horizontal $\xi_c/d_{monolayer}$ is in logarithm scale. The three straight lines clearly shows an empirical law $-\Delta T_c/T_c^{bulk} \sim \log(\xi_c/d_{monolayer})$ for the thin flakes from monolayer to 3-layer. Since the $T_c$ of 4-layer Cs-12442 is almost the same as the bulk single crystal, this empirical law does not hold for thick flakes with 4 layers and above. For $CaKFe_4As_4$, the $T_c$ of 3-layer flake is reduced by about 46%, and it is hard to get thinner flakes by cleaving. Therefore, 1144 system does not fit in this empirical law, which may be due to its much smaller anisotropy. This is also the case for FeSe, the simplest compound among iron-based superconductors. The anisotropy of FeSe is even smaller (about 3 ~ 4)[42], and the $\xi_c$ of FeSe is about 6 times the thickness of one monolayer[43]. The $T_c$ reduction in FeSe thin flakes is even faster than in the 1144 system[21,44]. Interestingly, for the FeSe thin films ($\geq$ 2L) grown on bilayer graphene, $T_c$ scales inversely with the thickness[45], which also does not fit the empirical law. Therefore, we conclude that the empirical law $-\Delta T_c/T_c^{bulk} \sim \log(\xi_c/d_{monolayer})$ is applicable to the thin flakes (≤ 3L) of quasi-2D superconductors with anisotropy roughly larger than $10^2$.

Note that there may be some other exceptions which does not obey above empirical law. For example, the $T_c$ of FeSe monolayer grown on $SrTiO_3$ substrate can be remarkably enhanced to over 60 K by electron doping and interfacial electron-phonon coupling[46-49]. Without those additional effects such as doping, substrate, and competing orders, we believe that the empirical law revealed above is universal for layered superconductors. Based on it, one may extrapolate the $T_c$ of a new monolayer superconductor. This will be important for the application quasi-2D superconductor thin flakes or heterostructures, with stack growth of $NbSe_2$\$MoS_2$ as an example[12].

## Methods

**Fabricating Cs-12442 thin flakes.** High-quality single crystals of CsCa$_2$Fe$_4$As$_4$F$_2$ were grown from CsAs flux[16]. Al$_2$O$_3$-assisted exfoliation method was applied to obtain atomically thin Cs-12442 flakes[19]. Al$_2$O$_3$ thin film with a thickness of about 60 ~ 100 nm was deposited by thermally evaporating Al under an oxygen pressure of 10$^{-2}$ Pa on the freshly prepared surface of the bulk Cs-12442 single crystal. Then the Al$_2$O$_3$ film along with pieces of Cs-12442 thin flakes separated from the bulk was picked up with a thermal release tape. The Al$_2$O$_3$/Cs-12442 stack was subsequently released onto a piece of polydimethylsiloxane (PDMS) with the Cs-12442 side in contact with the PDMS surface upon heating. Next the PDMS was stamped onto a clean sapphire substrate and was peeled away, leaving the Al$_2$O$_3$ film covered with Cs-12442 thin flakes on the substrate. The thickness of the flakes was determined by an atomic force microscope, and then we correlated the topography image with optical contrast. We used transmission mode to obtain the optical image of the ultrathin flakes, and the optical transmission is defined as $G^T_{sample}/G^T_{substrate}$, where $G^T_{sample}$ and $G^T_{substrate}$ are the intensities of the transmission (T) through the sample and substrate, respectively, in the green channel of the image. The layer-number-dependent optical transmission is well described by the Beer-Lambert law (as shown in Fig. 1d), which enables us to determine the layer number quickly and reliably.

**Electrical transport measurements on Cs-12442.** Two methods were applied to fabricate electrical contacts to Cs-12442 ultrathin flakes: (1) indium cold welding and (2) direct metal deposition through a stencil mask. The indium cold welding method was used for most samples since there is no thermal perturbation during the fabrication of electrodes. The direct metal deposition method was applied for the samples in the gate-tuning experiment, and in this case, the evaporated Cr/Au electrodes have a good chemical inertness during the electrochemical gating. For the indium cold welding method, an indium thin foil was first cut into small stripes with a sharp vertex angle. Then a PDMS stamp was used to hold these stripes, with which we align the Cs-12442 ultrathin flakes under an optical microscope. We then pressed the indium stripes onto the sample to form electrical contacts, before lifting off the PDMS stamp. Since Cs-12442 thin flakes are very sensitive to the air, all the device-fabrication processes were performed in an argon-filled glove box to avoid sample degradation. The resistance measurements of the thin flakes were performed in a 4 K cryocooler equipped with electrical transport systems (Multi-Fields & Group) inside the glove box, and in a physical properties measurement system (PPMS, Quantum Design) when magnetic field was needed.

**Gating devices for Cs-12442 ultrathin flakes.** The Cs-12442 ultrathin flakes were immersed in the ionic liquid *N*,*N*-diethyl-*N*-(2-methoxyethyl)-*N*-methylammonium bis(trifluoromethylsulfonyl) imide, and a large area gold electrode was used as the gating electrode, which was also immersed in the ionic liquid. For the gating process, the sample was first warmed up to 220 K, then the $V_g$ was applied to 3 V with the speed of 0.1 V/min, and held for $t_g$ at least 1 hour. $V_g$, $T_g$ and $t_g$ were further increased in the following gating procedures based on this initial condition. Finally, the sample was cooled down with the speed of 1 K/min, in order to avoid the accidental fragmentation during the freezing of ionic liquid.

Single crystals of $CaKFe_4As_4$ were grown using the self-flux method. The fabrication and measurement methods of $CaKFe_4As_4$ thin flakes are the same as Cs-12442.

## Data availability

The data that support the findings of this study are available from the corresponding authors upon reasonable request.

## Acknowledgements

We thank Tianping Ying and Yi Zhou for helpful discussions. This work is supported by the Natural Science Foundation of China (Grant No. 12174064), the Shanghai Municipal Science and Technology Major Project (Grant No. 2019SHZDZX01), and the National Key Research and Development Program of China (Grant No. 2022YFA1403202).

## Author Contributions

S.Y.L. conceived the idea and designed the experiments. K.M. performed the sample fabrication and electrical transport measurements. B.Z.L. and G.H.C. synthesized the single crystal samples. B.Q.S., X.M.K., S.C.H., X.B.J., Y.Y.W. and J.Y.N. assisted in the experimental setup. X.M.K. and X.F.Y. offered beneficial suggestions on the experiments and assisted in drawing the illustrations. K.M., X.Z., B.Q.S. and S.Y.L. analyzed the experimental data and wrote this paper with comments from all authors. K.M., X.Z., B.Q.S., B.Z.L. contributed equally to this work.

## Competing interests

The authors declare no competing interests.

## Additional Information

**Supplementary information** is available for this paper at URL inserted when

Figure 1

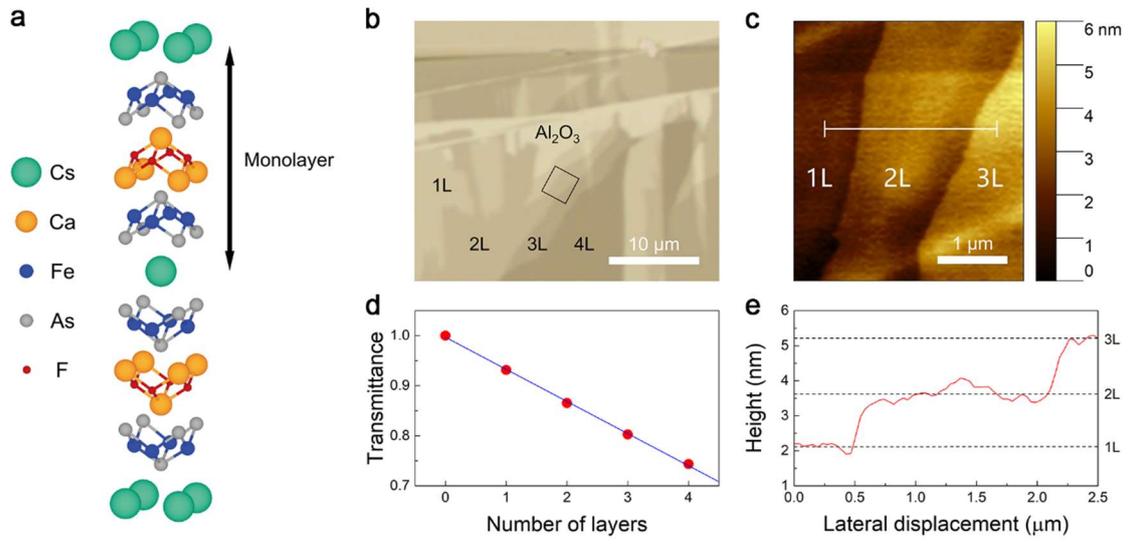

Figure 2

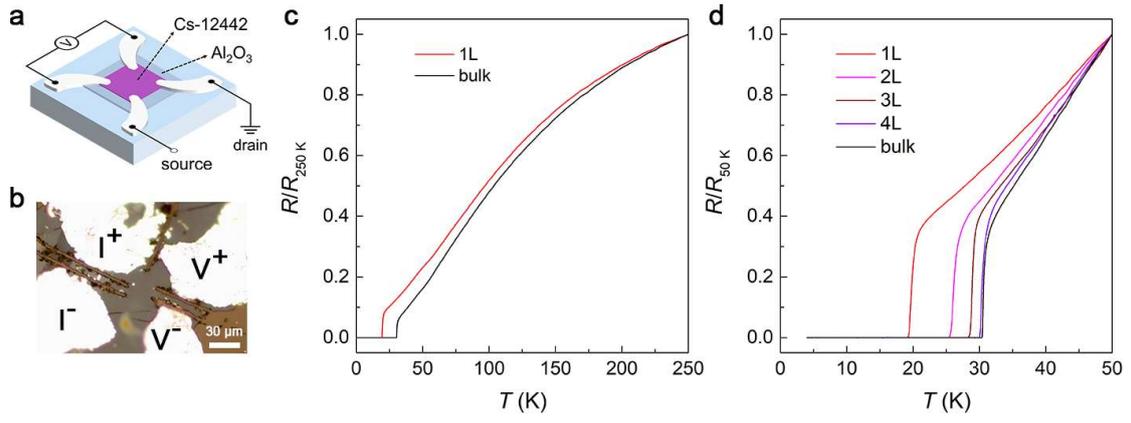

Figure 3

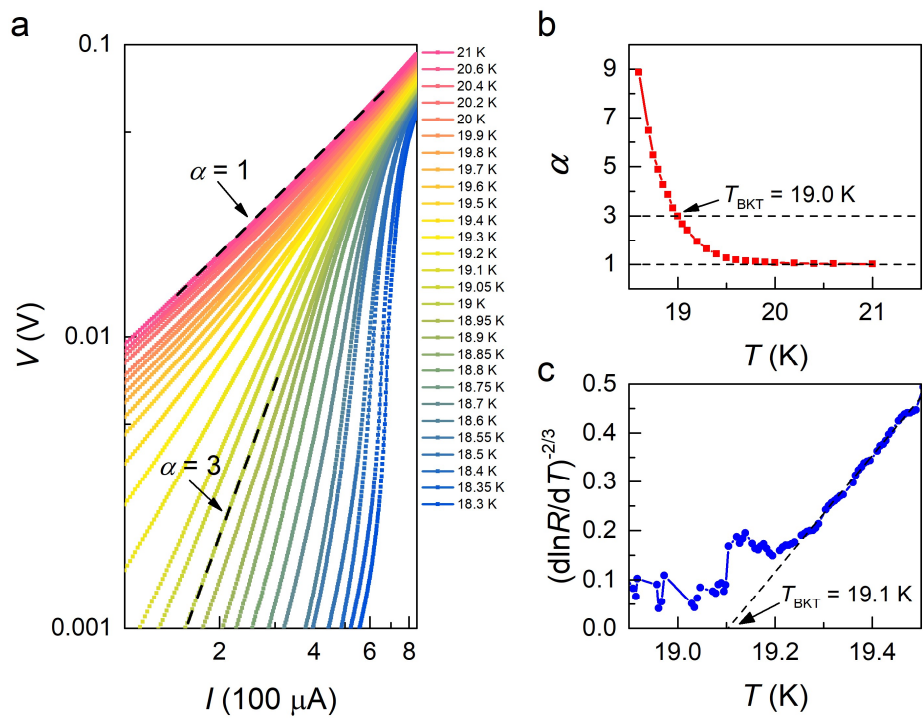

Figure 4

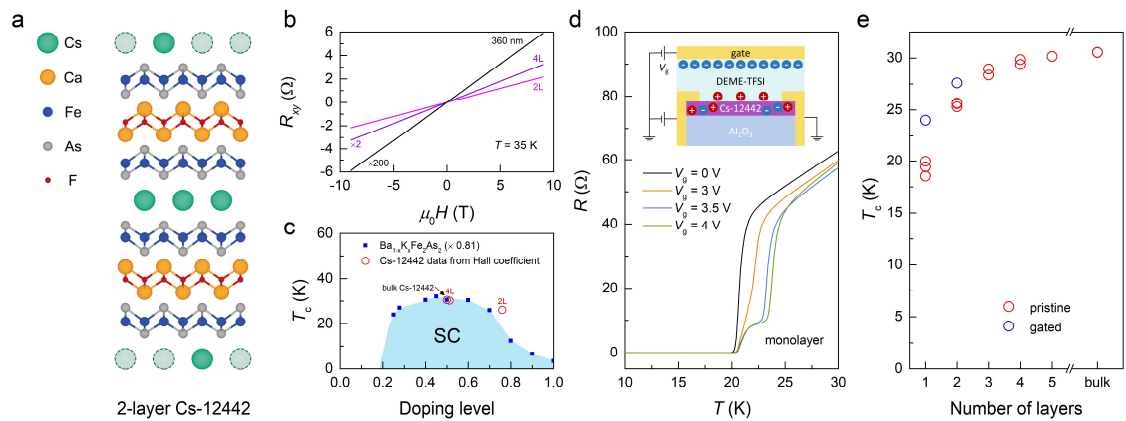

Figure 5

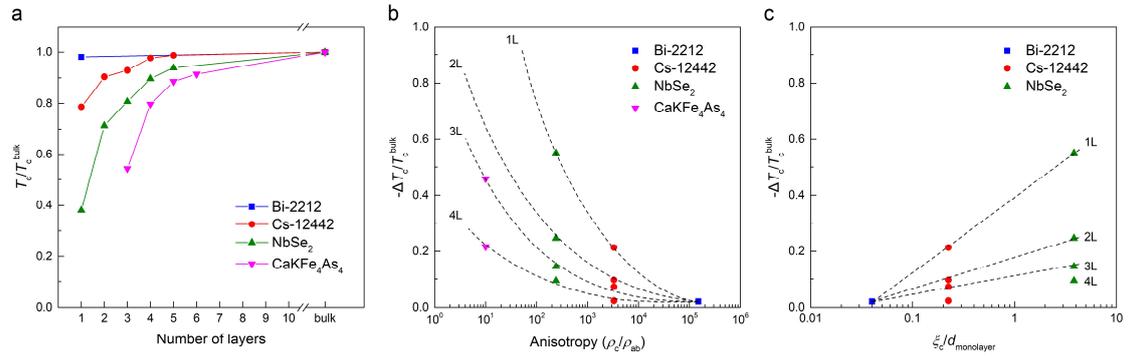

**Table 1** | Anisotropy ($\rho_c/\rho_{ab}$), *c*-axis coherence length ($\xi_c$), $\xi_c/d_{monolayer}$ ($d_{monolayer}$ is the thickness of monolayer), $T_c$, reduction of $T_c$ ($-\Delta T_c/T_c^{bulk}$) for Bi-2212, Cs-12442, NbSe$_2$, CaKFe$_4$As$_4$ and their corresponding monolayers (3-layer for CaKFe$_4$As$_4$).

|  | $\rho_c/\rho_{ab}$ | $\xi_c$ | $\xi_c/d_{monolayer}$ | $T_c$ | $-\Delta T_c/T_c^{bulk}$ |
|---|---|---|---|---|---|
| **Bi-2212** | $1.5 \times 10^5$ (ref.[7]) | ≤ 0.6 Å (ref.[9]) | < 0.04 | 88.6 K (ref.[6]) |  |
| **Bi-2212 monolayer** |  |  |  | 86.7 K (ref.[6]) | 2% |
| **Cs-12442** | 3150 (ref.[16]) | 3.6 Å (ref.[16]) | 0.225 | 30.5 K |  |
| **Cs-12442 monolayer** |  |  |  | 24.0 K | 21% |
| **NbSe$_2$** | 240 (ref.[10]) | 23 Å (ref.[11]) | 3.84 | 6.9 K (ref.[5]) |  |
| **NbSe$_2$ monolayer** |  |  |  | 3.1 K (ref.[5]) | 55% |
| **CaKFe$_4$As$_4$** | ~ 10 (ref.[17]) | 5.8 Å (ref.[17]) | 0.91 | 35.2 K |  |
| **CaKFe$_4$As$_4$ 3-layer** |  |  |  | 19.0 K | 46% |

**Figure captions**

**Figure 1 | Characterization of Cs-12442 ultrathin flakes. a** Atomic structure of Cs-12442. 'Monolayer' refers to a Cs-Fe$_2$As$_2$-Ca$_2$F$_2$-Fe$_2$As$_2$-Cs half-unit cell along the out-of-plane *c*-axis direction. **b** Optical image of typical few-layer Cs-12442 flakes exfoliated on top of a thermally evaporated Al$_2$O$_3$ thin film (thickness ~ 100 nm). The Al$_2$O$_3$ film is supported on a quartz substrate. The image is taken in transmission mode. Number of layers is labeled on selected flakes. **c** Atomic force microscopy (AFM) image of the same flake shown in **b** (region marked by the black square). **d** Transmittance as a function of the number of layers. The transmittance (red dots) follows the Beer-Lambert law (blue line). **e** Cross-sectional profile of the AFM topography along the white line in **c**, the step is 1.6 nm in height which corresponds to the thickness of the Cs-12442 monolayer.

**Figure 2 | Fabricating electrodes and electrical transport measurement of Cs-12442 ultrathin flakes. a** A schematic illustration of the Cs-12442 ultrathin flake device. **b** The optical image of a monolayer Cs-12442 device. The sample is contacted by pressed indium electrodes. The measurement set-up is also marked. **c** Temperature dependence of the normalized resistance $R/R_{250\text{ K}}$ for a bulk single crystal and a monolayer Cs-12442. **d** Details of the normalized resistance $R/R_{50\text{ K}}$ as a function of temperature for a bulk single crystal and various ultrathin flakes with layer numbers from 4 to 1. The reduction of $T_\text{c}$ is evident when approaching the monolayer limit.

**Figure 3 | The BKT transition in monolayer Cs-12442. a** The *V-I* relationship at different temperatures in superconducting regime for monolayer Cs-12442. The curves are plotted in logarithmic scale. The two black dashed lines refer to $V \sim I$ and $V \sim I^3$, respectively. **b** Variation of exponent $\alpha$ as a function of temperature, extracting from the power-law fittings in **a**, showing that $T_\text{BKT} = 19.0$ K. **c** The *R-T* curve plotted with the $[\text{dln}(R)/\text{d}T]^{-2/3}$ scale. The dashed line shows the fitting to the Halperin-Nelson formula $R(T) = R_0\exp[-b/(T - T_\text{BKT})^{-1/2}]$, giving $T_\text{BKT} = 19.1$ K.

**Figure 4 | Doping effect and tunable superconductivity of Cs-12442 ultrathin flakes. a** A cross section schematic of 2-layer Cs-12442. The deficiency of surface Cs atoms is marked by dashed green circles. **b** Raw Hall data of 2-layer, 4-layer and 360-nm Cs-12442 flakes at $T = 35$ K. **c** $T_c$ as a function of hole doping level. The doping level is determined by the value $x$ in $Ba_{1-x}K_xFe_2As_2$ (ref.[28]). Here, we consider the 360-nm flake to have the same doping level as the bulk crystal, corresponding to $x = 0.5$ in the phase diagram. For the 2-layer and 4-layer flakes, the doping level $x$ is obtained from the equation $x = (n_{\text{thin flake}}/n_{\text{bulk}}) \times 0.5$, $n$ stands for the hole concentration. **d** Temperature dependence of the resistance for a monolayer Cs-12442 at different gating voltage, $V_g$. The inset is a schematic of an ionic liquid gating device. Here, we show the penetration of small-sized cations into the sample, which induces a negative charge as a result. **e** $T_c$ as a function of layer numbers for Cs-12442. Here, $T_c$ is defined as the temperature corresponding to the maximum value of the derivative of the $R$-$T$ curve during the superconducting transition. The red circles represent pristine samples, and the blue circles represent gate-tuned samples with highest $T_c$.

**Figure 5 | Reduction of $T_c$ near the 2D limit as a function of anisotropy and coherence length. a** Layer dependence of $T_c/T_c^{\text{bulk}}$ for Bi-2212, Cs-12442, CaKFe$_4$As$_4$, and NbSe$_2$. **b** Reduction of $T_c$ (defined as $-\Delta T_c/T_c^{\text{bulk}}$) as a function of anisotropy ($\rho_c/\rho_{ab}$) for these materials with various number of layers. The dashed curves represent the general trand. The data of anisotropy are from ref.[7,10,16,17]. **c** $-\Delta T_c/T_c^{\text{bulk}}$ as a function of the ratio of coherence length ($\xi_c$) to monolayer thickness ($d_{\text{monolayer}}$) for Bi-2212, Cs-12442, and NbSe$_2$. The three straight lines clearly shows an empirical law $-\Delta T_c/T_c^{\text{bulk}} \sim \log(\xi_c/d_{\text{monolayer}})$ for the thin flakes from monolayer to 3-layer. The data of coherence length are from ref.[9,11,16].

# Supplementary Information for

# "Layer-dependent superconductivity in iron-based superconductors"

Ke Meng[1,2,*], Xu Zhang[1,*], Boqin Song[1,*], Baizhuo Li[3,*], Xiangming Kong[1], Sicheng Huang[1], Xiaofan Yang[1], Xiaobo Jin[1], Yiyuan Wu[1], Jiaying Nie[1], Guanghan Cao[3,4,✉], and Shiyan Li[1,2,4,✉]

[1]*State Key Laboratory of Surface Physics, Department of Physics, Fudan University, Shanghai 200438, China*

[2]*Shanghai Research Center for Quantum Sciences, Shanghai 201315, China*

[3]*School of Physics, Zhejiang University, Hangzhou 310058, China*

[4]*Collaborative Innovation Center of Advanced Microstructures, Nanjing 210093, China*

Corresponding author. Email: shiyan_li@fudan.edu.cn (S.Y.L.); ghcao@zju.edu.cn (G.H.C.)

# Supplementary Note 1: Characterization of the number of layers and superconductivity for Cs-12442 ultrathin flakes

Here, we present a detailed characterization of the layer number and superconductivity for typical 1 ~ 4-layer Cs-12442, as shown in Supplementary Fig. S1. Due to the relatively thick monolayer (~ 1.6 nm) in this system, a clear difference in optical contrast is observed for different numbers of layers. Furthermore, for ultrathin flakes, the transmittance follows the Beer-Lambert law, which allows us to reliably identify the layer numbers. The superconductivity of 1 ~ 4-layer Cs-12442 was measured by four-probe electrical transport and the results are shown in Supplementary Fig. S1. All samples exhibit a sharp superconducting transition with different $T_c$.

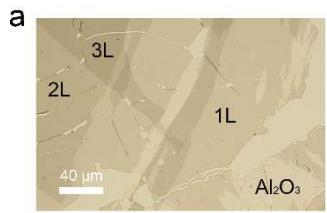
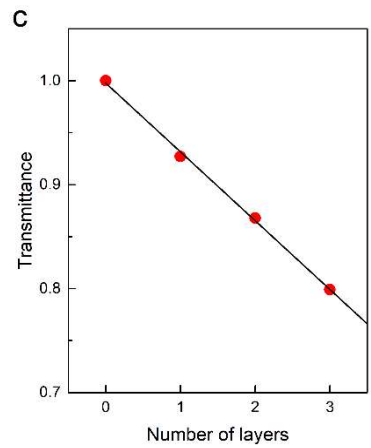
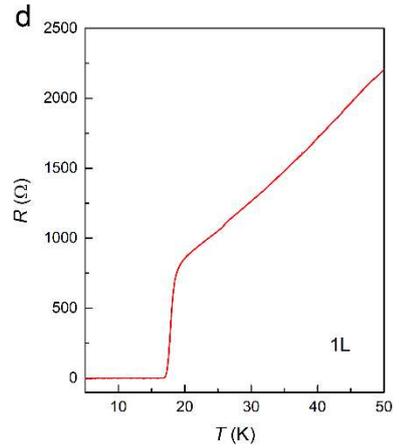
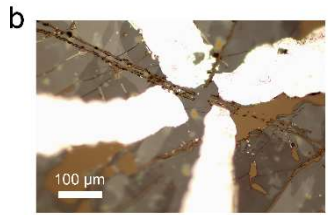
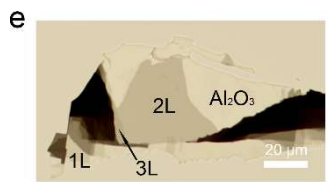
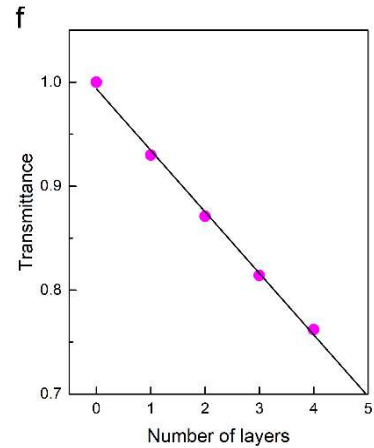
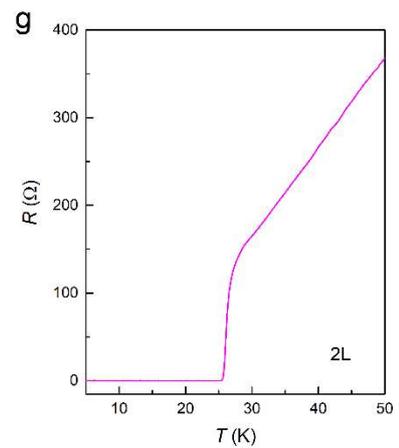
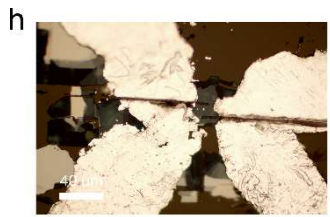
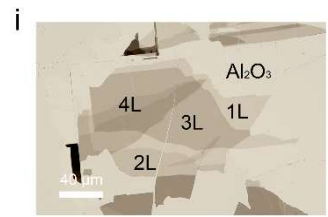
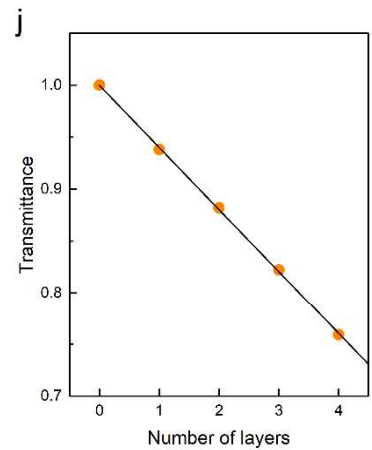
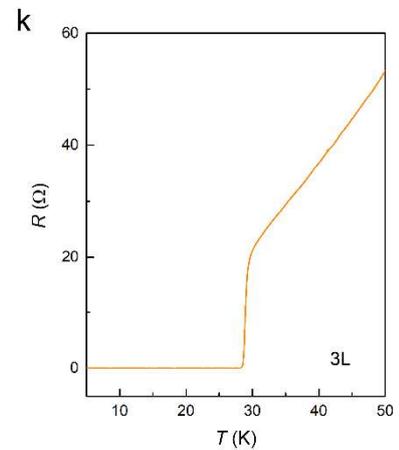
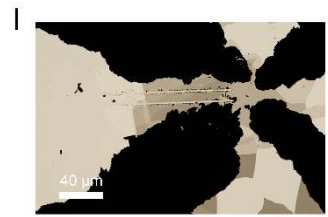
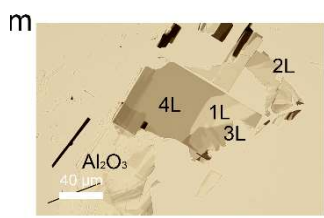
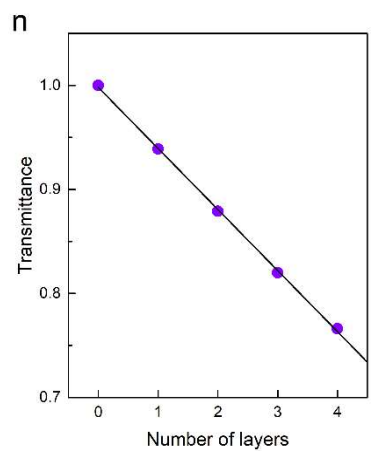
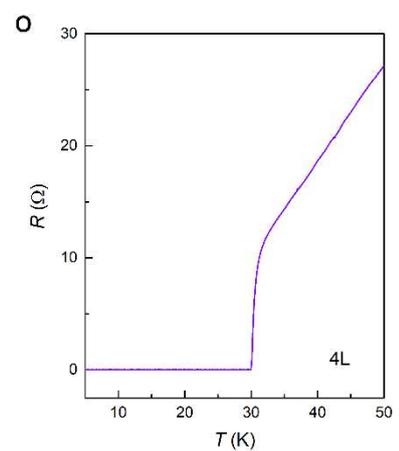
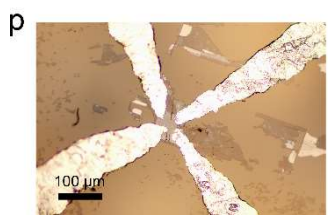

**Supplementary Figure S1 | Characterization and superconductivity of monolayer, 2-layer, 3-layer, and 4-layer Cs-12442. a, e, i, m** Optical images of 1 ~ 4-layers Cs-12442 flakes exfoliated on top of a thermally evaporated $Al_2O_3$ thin film. The images are taken in transmission mode. The number of layers is labeled on selected flakes. **b, h, l, p** Optical images of the 1 ~ 4-layer Cs-12442 device. Samples are contacted by pressed indium electrodes. **c, f, j, n** Transmittance as a function of the number of layers for thin flakes as shown in **a, e, i, m**. All the transmittance follows the Beer-Lambert law (black line). **d, g, k, o** Temperature dependence of the resistance for 1 ~ 4 layers Cs-12442 samples.

# Supplementary Note 2: Influence from degradation on the ultrathin-flake device

To investigate degradation effects on ultrathin-flake device, $T_c$ is measured as a function of exposure time in the glove box for the Cs-12442 monolayer, as shown in Supplementary Fig. S2. Here, our electrical transport measurement setup is directly exposed to the glove box atmosphere and the exposure conditions are the same as that during the sample fabrication process (room temperature and water and oxygen content below 0.6 ppm). The fabrication of a sample, which involves exfoliation, transfer, and pressure welding electrodes, takes less than three hours on average. Our study aims to examine whether there is a significant change in $T_c$ over a much longer period of exposure time in the glove box, particularly in the case of the monolayer sample. Our results show that the $T_c$ of monolayer Cs-12442 decreases by approximately 1 K following a total exposure time of up to 8 hours. Therefore, the influence of degradation during sample fabrication in the glove box is negligible, due to the robust protection provided by the argon atmosphere.

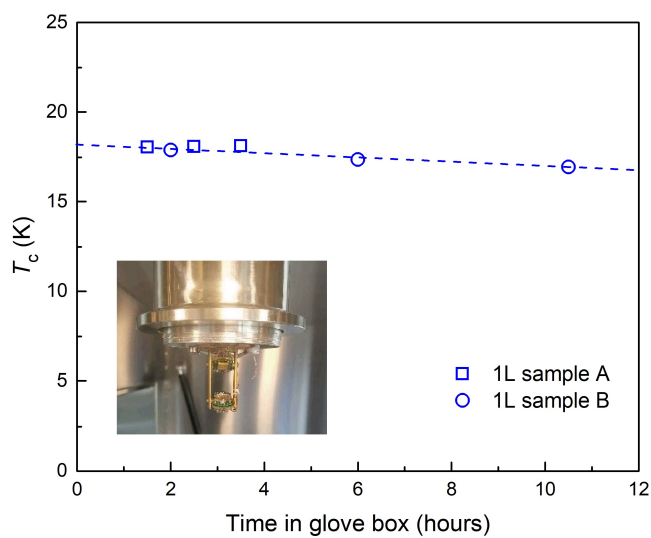

**Supplementary Figure S2 | Influence of degradation on the ultrathin-flake device.** $T_c$ as a function of exposure time in glove box for monolayer Cs-12442 is displayed. Dashed line marks the tendency for the change of $T_c$. The inside image is the set up for the exposure time study.

## Supplementary Note 3: Gate-tunable superconductivity in Cs-12442 ultrathin flakes.

The ionic liquid gating technique is used to modulate the doping level of Cs-12442 ultrathin flakes. In our experiment, several parameters affect the gating process together: gating voltage ($V_g$), gating temperature ($T_g$), and gating duration time ($t_g$). By adjusting $V_g$, $T_g$, and $t_g$, the superconductivity of the Cs-12442 ultrathin flakes can be efficiently tuned. The detailed results of the gate-tunable superconductivity of monolayer and 2-layer Cs-12442 are presented in Supplementary Fig. S3 and Supplementary Fig. S4. The gate-tuning experiment for the monolayer sample contains 14 rounds, and the enhancement of $T_c$ tends to saturate at the end of the gating. The curves show two superconducting transitions. The lower $T_c$ does not change much with gating, which may be due to the fact that small-sized H$^+$ ions do not enter into parts of the sample, likely the area under that electrodes. Here, $T_c$ is defined as the temperature at which the differential of the $R$-$T$ curve reaches the maximum, as shown in Supplementary Fig. S3c and Supplementary Fig. S4b.

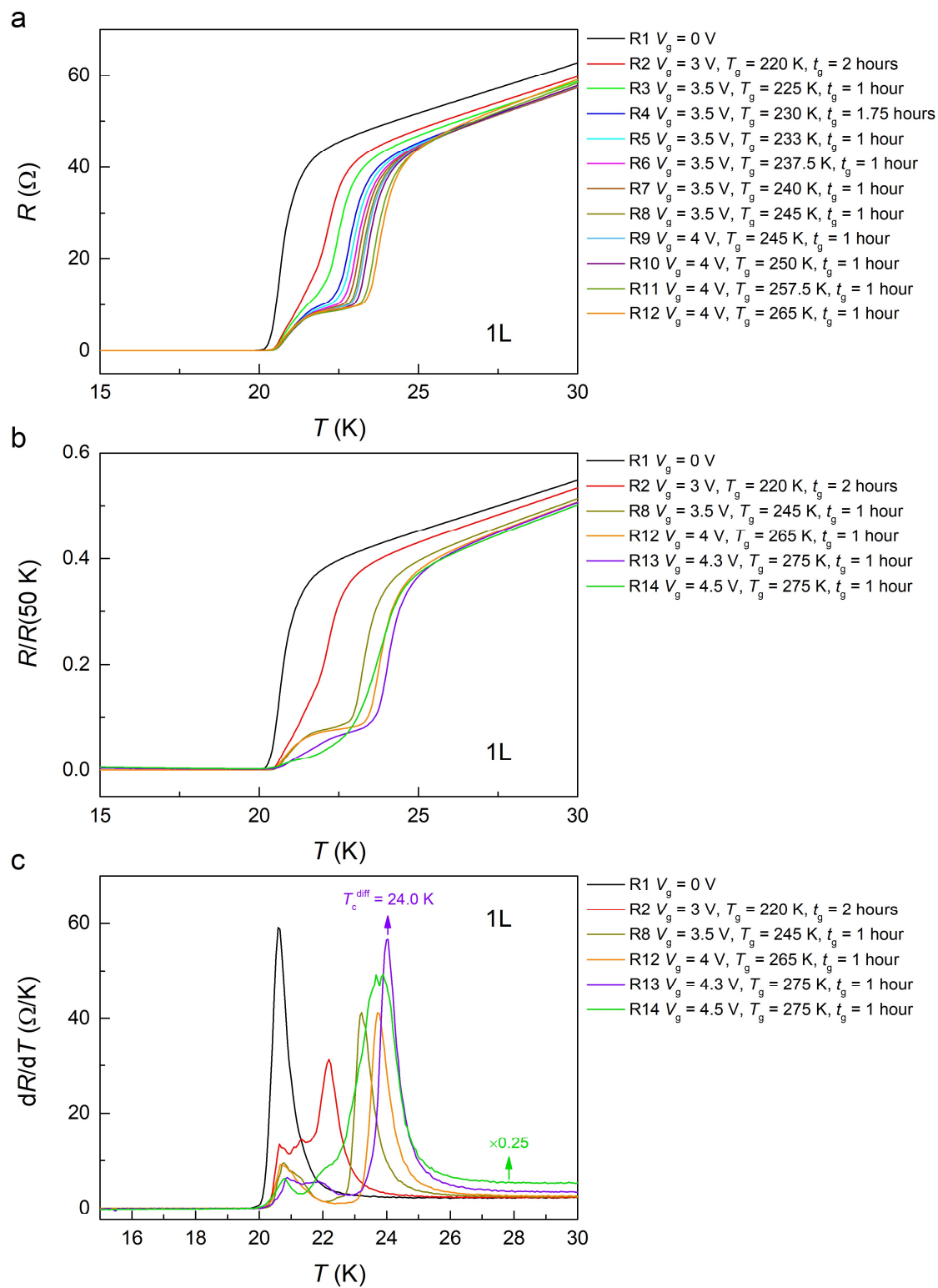

**Supplementary Figure S3 | Gate-tunable superconductivity in monolayer Cs-12442. a** Temperature dependence of the resistance for a monolayer Cs-12442 for successive gating sequences. Detailed gating conditions of gating voltage ($V_g$), gating temperature ($T_g$) and gating duration time ($t_g$) are displayed. **b** Temperature dependence

of the normalized resistance $R/R(50\ \text{K})$ for the same monolayer in **a**. **c** First derivative of resistance ($dR/dT$) as a function of temperature for the same monolayer.

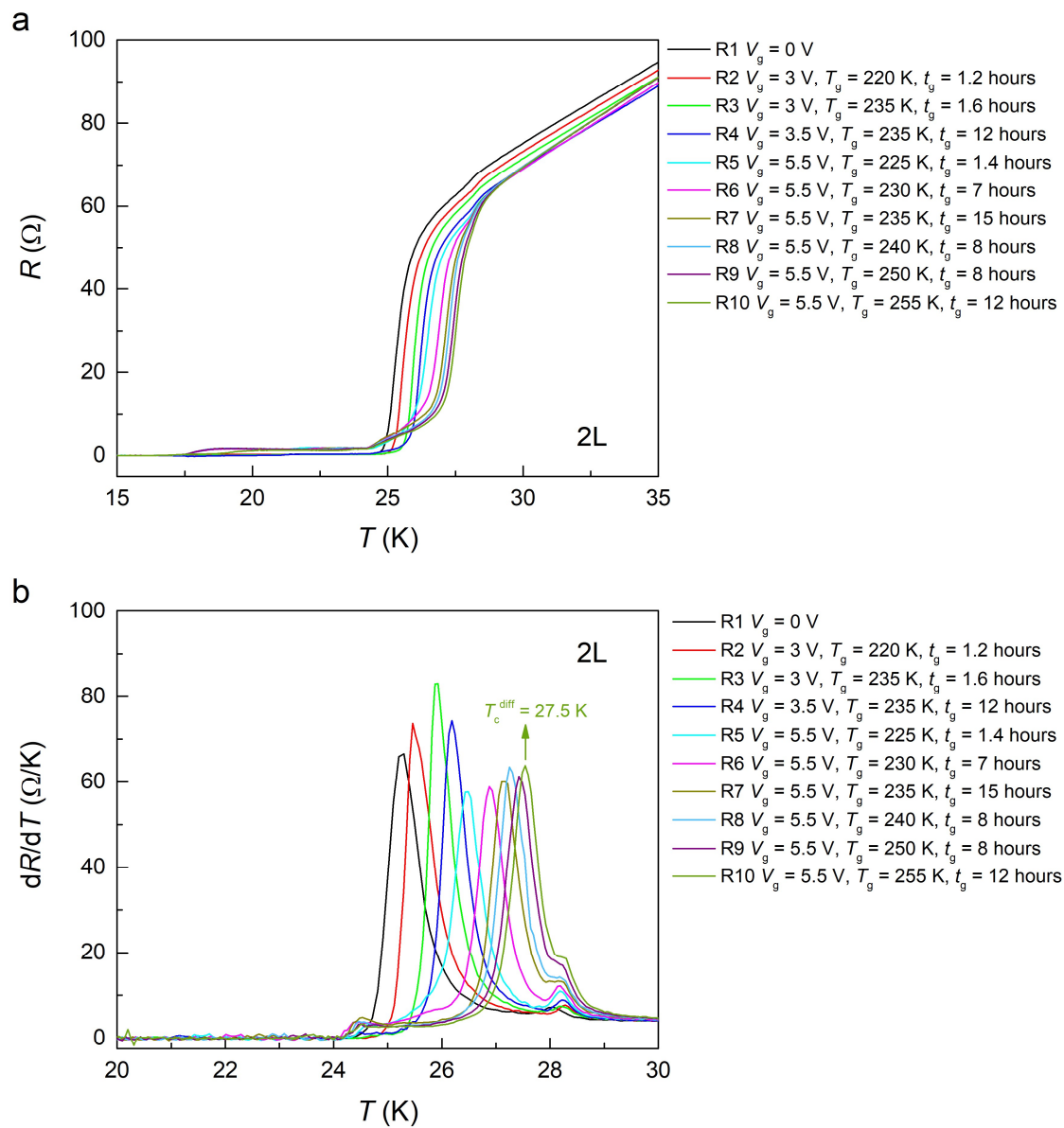

**Supplementary Figure S4 | Gate-tunable superconductivity in 2-layer Cs-12442. a** Temperature dependence of the resistance for the 2-layer Cs-12442 for successive gating sequences. Detailed gating conditions of gating voltage ($V_g$), gating temperature ($T_g$) and gating duration time ($t_g$) are displayed. **b** First derivative of resistance ($dR/dT$) as a function of temperature for the same sample in **a**.

## Supplementary Note 4: Layer-dependent superconductivity of CaKFe$_4$As$_4$.

We have also been successful in fabricating ultrathin flakes of CaKFe$_4$As$_4$, as demonstrated in Supplementary Fig. S5. In contrast to Cs-12442, it is more challenging to obtain large-area ultrathin flakes of CaKFe$_4$As$_4$, which may be due to its stronger interlayer coupling. In our experiments, regions of different thicknesses are often coexisting between the V$^+$ and V$^-$ electrodes, sometimes resulting in more than one superconducting transition in the $R$ - $T$ curve. Meanwhile, the superconducting transition width of the ultrathin flakes is quite broad. Therefore, the $T_c$ is defined at the onset temperature of the superconducting transition ($T_c^{onset}$). Note that we have not succeeded in preparing a monolayer CaKFe$_4$As$_4$.

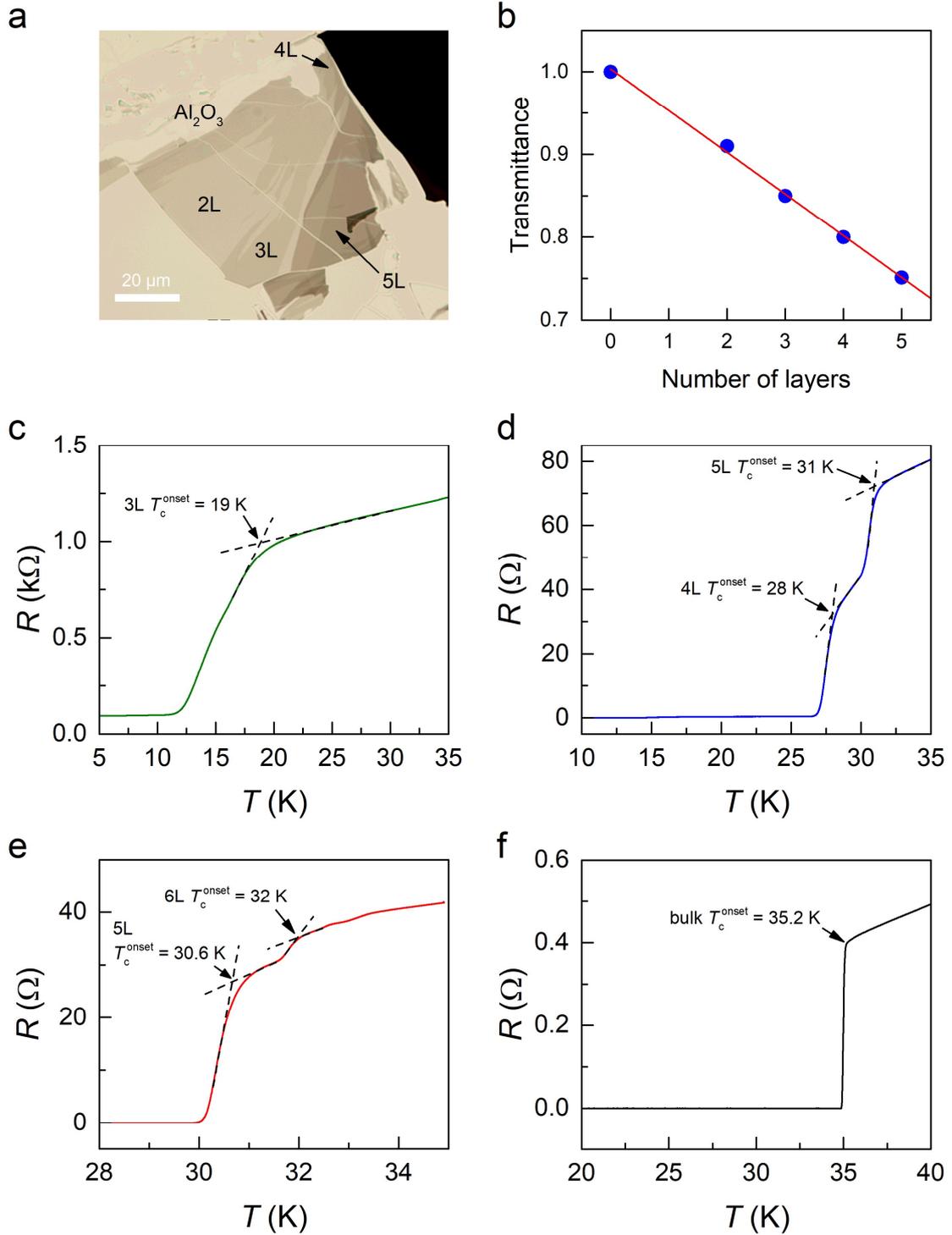

**Supplementary Figure S5 | Layer-dependent superconductivity in CaKFe₄As₄. a** A typical optical image of CaKFe₄As₄ ultrathin flakes exfoliated on top of a thermally evaporated Al₂O₃ thin film. The image is taken in transmission mode. Number of layers is labeled on selected flakes. **b** Transmittance as a function of the number of layers for thin flakes shown in **a**. All the transmittance follows the Beer-Lambert law (red line). **c, d, e, f** Temperature dependence of the resistance for 3-, 4-, 5-, 6-layer and bulk

CaKFe$_4$As$_4$. In **d** and **e**, both samples show more than one superconducting transition, due to the fact that the measured region between the electrodes contains different layers. The $T_c$ corresponding to various number of layers is marked. Here, the $T_c$ is defined by $T_c^{onset}$, which is determined from the cross point of two dash lines.